\documentclass[
onlinecite, reprint, 
superscriptaddress,
amsmath,amssymb,
aps,prl
]{revtex4-2}

\usepackage{makecell}
\usepackage{graphicx}% Include figure files
\usepackage{dcolumn}% Align table columns on decimal point
\usepackage{bm}% bold math
\usepackage{longtable}
\usepackage{physics}
\usepackage{amsmath}
\usepackage{multirow}
\usepackage[table,dvipsnames]{xcolor}
\usepackage{colortbl}
\usepackage{booktabs}
\usepackage{soul}
\usepackage{comment}
\usepackage{xcolor}
\usepackage[
    colorlinks=true,
    linkcolor=blue,
    citecolor=blue,
    urlcolor=blue
]{hyperref}

\begin{document}
\title{Zeeman Spin Photocurrents without Spin Textures}

\author{Yihua Xiao}
\affiliation{Key Lab of advanced optoelectronic quantum architecture and measurement (MOE),
Beijing Key Laboratory of Quantum Matter State Control and Ultra-Precision Measurement Technology,
and School of Physics, Beijing Institute of Technology, Beijing 100081, China}

\author{Ruizhi Dong}
\affiliation{Key Lab of advanced optoelectronic quantum architecture and measurement (MOE),
Beijing Key Laboratory of Quantum Matter State Control and Ultra-Precision Measurement Technology,
and School of Physics, Beijing Institute of Technology, Beijing 100081, China}

\author{Ruixiang Fei}
\email{rfei@bit.edu.cn}
\affiliation{Key Lab of advanced optoelectronic quantum architecture and measurement (MOE),
Beijing Key Laboratory of Quantum Matter State Control and Ultra-Precision Measurement Technology,
and School of Physics, Beijing Institute of Technology, Beijing 100081, China}

\begin{abstract}
Inversion symmetry forbids bulk second-order spin photocurrents driven solely by electric-dipole coupling. We show that interference between the electric-dipole and spin Zeeman couplings of light enables Zeeman spin-shift and spin-injection currents in centrosymmetric magnets. Our microscopic theory and systematic symmetry analysis establish the allowed responses and polarization selection rules for both spin-degenerate and spin-split antiferromagnets, with and without spin--orbit coupling (SOC). In collinear magnets without SOC, spin injection vanishes, whereas a pure spin-shift current survives independently of the carrier scattering time. This response arises from transverse electronic spin coherences generated by the optical magnetic field, while the ordered moments remain collinear and fixed. Model analyses validate the mechanism, and first-principles calculations for $\mathrm{MnF_2}$ predict experimentally accessible spin photocurrents. These findings establish a general framework for optical spin-current generation in centrosymmetric magnets, including a SOC-free pure-spin mechanism particularly relevant to altermagnets.
\end{abstract}
%\pacs{Valid PACS appear here}
%PACS, the Physics Classification Scheme.
%\keywords{Suggested keywords} $Use show keys class option if keyword display desired.
\maketitle

\textcolor{blue}{\textbf{Introduction}}---
Optical generation of spin currents offers ultrafast control of angular-momentum transport~\cite{Bhat2000,Stevens2003,Kirilyuk2010}. Within the electric-dipole approximation, however, bulk
second-order charge and spin photocurrents both require
broken inversion symmetry~\cite{sipe2000,Glass1974,vonBaltz1981,Fei2020,Lihm2022,Young2012,Morimoto2016,deJuan2017,Habara2023}. 
This requirement follows from the opposite inversion parities of the spin-current operator and the quadratic optical electric-field product. Contributions from inversion-related momenta therefore cancel in centrosymmetric crystals, forbidding conventional spin shift and injection currents.

Collinear magnets without SOC are attractive for spin transport because they retain a common spin-quantization axis and avoid SOC-induced relaxation channels~\cite{Elliott1954,Zutic2004,Denisov2024,GonzalezHernandez2021,Sheoran2026,Chen2025}. Altermagnets provide a natural platform, combining spin-split electronic bands without SOC with collinear magnetic order and zero net magnetization~\cite{Hayami2019,Smejkal2020,Yuan2020}. Harnessing these advantages under optical excitation calls for spin-current generation mechanisms that require neither SOC nor noncollinear spin textures in real or momentum space.
However, representative altermagnetic systems such as
RuO$_2$, CrSb, MnF$_2$, and
$\alpha$-MnTe~\cite{Krempasky2024,Lee2024,Zhou2024,Wan2025,Yuan2020,Roig2024,Zhou2025,Liu2022,Smejkal2022,Ding2024,Ohlendorf2026} are centrosymmetric,
precluding bulk second-order spin photocurrents within
the electric-dipole approximation~\cite{Yoshida2026,Lihm2022,Kim2017,Mokrousov2021,Sivianes2025,Hsu2025,Ishizuka2019,Xiao2021}.

Recent work has identified Zeeman shift and injection charge currents arising from electric-dipole–Zeeman interference, demonstrating that the magnetic component of light can circumvent this inversion constraint~\cite{Dong2026zeeman,Xiang2026}. 
These charge responses are determined in part by Zeeman
quantum geometry involving the Pauli spin operators~\cite{Freimuth2014,Gao2018,Dong2020,Xiang2025,Hattori2025}
and thus require spin mixing induced by SOC or noncollinear
magnetic order.
The corresponding spin photocurrents have yet to be
systematically investigated. In particular, whether the
same interference mechanism can generate spin currents
in collinear magnets without SOC remains an open question.

In this work, we develop a microscopic theory of Zeeman shift and injection spin currents from
electric-dipole–Zeeman interference, establish their symmetry and polarization selection rules, and predict detectable responses in materials with SOC using model and
first-principles calculations. Crucially, Zeeman shift spin currents persist in 
collinear altermagnets without SOC, where the optical magnetic field generates the 
spin coherence needed for a pure transverse spin current while the ordered 
moments remain fixed.

\textcolor{blue}{\textbf{Microscopic mechanism of Zeeman shift and injection spin currents}}---
We consider electrons in spatially uniform optical fields,
\begin{equation}
  h(t)=h_0-eE^b(t)r^b-\lambda_Z B^\beta(t)\sigma^\beta,
  \label{eq:spin_ham}
\end{equation}
where $h_0$ is the equilibrium Hamiltonian, $e$ the signed electron
charge, and $\lambda_Z=g\mu_B/2$ the spin Zeeman coupling strength,
with $\sigma^\beta$ a Pauli operator.

%We formulate the spin photocurrent response of an equilibrium insulator using Bloch eigenstates $|u_{n\mathbf{k}}\rangle$. We define the transition frequencies $\omega_{nm}=(\epsilon_n-\epsilon_m)/\hbar$, the equilibrium occupation differences $f_{nm}=f_n^0-f_m^0$, and the Berry connection $\mathcal A^b_{nm}=i\langle u_n|\partial_b u_m\rangle$.
%For Bloch eigenstates $|u_{n\mathbf k}\rangle$, we define $\omega_{nm}=(\epsilon_n-\epsilon_m)/\hbar$, $f_{nm}=f_n^0-f_m^0$, and $\mathcal A^b_{nm}=i\langle u_n|\partial_bu_m\rangle$. The interband position matrix is $r^b_{nm}=\mathcal A^b_{nm}$ for $n\ne m$.
%, with $r^b_{nn}=0$. Covariant derivatives are $O_{nm;b}=\partial_bO_{nm}-i(A_n^b-A_m^b)O_{nm}$, where $A_n^b=\mathcal A^b_{nn}$ and $\partial_b=\partial/\partial k_b$.

%For an equilibrium insulator with Bloch states $|u_{n\mathbf{k}}\rangle$, we define $\omega_{nm}=(\epsilon_n-\epsilon_m)/\hbar$, $f_{nm}=f_n^0-f_m^0$, and $\mathcal A^b_{nm}=i\langle u_n|\partial_bu_m\rangle$. The interband position matrix is $r^b_{nm}=\mathcal A^b_{nm}$ for $n\ne m$, and the covariant derivative is $O_{nm;b}=\partial_bO_{nm}-i(A_n^b-A_m^b)O_{nm}$, with $A_n^b=\mathcal A^b_{nn}$ and $\partial_b=\partial/\partial k_b$.

We use the conventional spin-current operator
\begin{equation}
  j_a^s=\frac e2\{v_0^a,\sigma^s\},\qquad
  v_0^a=\frac{[r^a,h_0]}{i\hbar},
  \label{eq:spin_operator}
\end{equation}
where $a$ and $s$ label the flow direction and spin polarization.
We use the spin-current operator directly, rather than the
polarization-operator approach of Sipe \textit{et al.}~\cite{sipe2000}
%and its Zeeman extension by
and Dong \textit{et al.}~\cite{Dong2026zeeman}. Its Bloch matrix elements are
%where $a$ and $s$ label the flow direction and spin polarization. It has no explicit optical-field dependence and reads in the Bloch basis
\begin{equation}
\begin{aligned}
  (j_a^s)_{nm}=\frac e2\Big[&
  (v_n^a+v_m^a)\sigma^s_{nm}\\
  &+i\sum_l\big(\omega_{nl}r^a_{nl}\sigma^s_{lm}
  +\omega_{lm}\sigma^s_{nl}r^a_{lm}\big)\Big],
\end{aligned}
  \label{eq:spin_matrix}
\end{equation}
where $v_n^a=\partial_a(\epsilon_n/\hbar)$,
$\omega_{nm}=(\epsilon_n-\epsilon_m)/\hbar$, and
$f_{nm}=f_n^0-f_m^0$ denotes the equilibrium occupation difference.

The second-quantized current and density matrix are
\begin{equation}
\hat J_a^s=\frac1V\sum_{nm}\int dk\,(j_a^s)_{nm}a_n^\dagger a_m,
\qquad \rho_{nm}=\langle a_m^\dagger a_n\rangle.
\label{eq:current-density-matrix}
\end{equation}
Thus the response follows from tracing the full current operator with
the field-induced density matrix.

\textit{Mixed dc response}---We use real optical fields
$E^b(t)=\mathcal E^b e^{-i\omega t}+\mathrm{c.c.}$
and
$B^\beta(t)=\mathcal B^\beta e^{-i\omega t}+\mathrm{c.c.}$,
with $\omega>0$.
For the insulating ground state,
$\rho^{(0)}_{nm}=f_n^0\delta_{nm}$ and $\partial_b f_n^0=0$.
The covariant derivative is
$O_{nm;b}=\partial_bO_{nm}-i(A_n^b-A_m^b)O_{nm}$, where
$A_n^b=i\langle u_n|\partial_bu_n\rangle$.

\begin{comment}
The Liouville equation $i\hbar\dot\rho=[h,\rho]$ yields the first-order response
\begin{equation}
\begin{aligned}
\rho^{(1),E}_{nm}(t)
&=-\frac{ef_{nm}r^b_{nm}}{\hbar}\\
&\quad\times\left[
\frac{\mathcal E^b e^{-i\omega t}}{\omega_{nm}-\omega-i\eta_E}
+\frac{\mathcal E^{b*}e^{i\omega t}}{\omega_{nm}+\omega-i\eta_E}
\right],\\
\rho^{(1),B}_{nm}(t)
&=-\frac{\lambda_Z f_{nm}\sigma^\beta_{nm}}{\hbar}\\
&\quad\times\left[
\frac{\mathcal B^\beta e^{-i\omega t}}{\omega_{nm}-\omega-i\eta_B}
+\frac{\mathcal B^{\beta*}e^{i\omega t}}{\omega_{nm}+\omega-i\eta_B}
\right].
\end{aligned}
\label{eq:rho1}
\end{equation}
\end{comment}

The Liouville equation $i\hbar\dot\rho=[h,\rho]$ gives
the first-order response
\begin{equation}
\begin{aligned}
\rho^{(1),E}_{nm}(t)
&=-\frac{ef_{nm}r^b_{nm}}{\hbar}
\sum_{\alpha=\pm1}
\frac{E^b(\alpha\omega)e^{-i\alpha\omega t}}
{\omega_{nm}-\alpha\omega-i\eta_E},\\
\rho^{(1),B}_{nm}(t)
&=-\frac{\lambda_Zf_{nm}\sigma^\beta_{nm}}{\hbar}
\sum_{\alpha=\pm1}
\frac{B^\beta(\alpha\omega)e^{-i\alpha\omega t}}
{\omega_{nm}-\alpha\omega-i\eta_B}.
\end{aligned}
\label{eq:rho1}
\end{equation}
Here $E^b(\omega)=\mathcal E^b$ and
$E^b(-\omega)=\mathcal E^{b*}$, and similarly for $B^\beta$.
Substituting these responses gives, at mixed second order,
\begin{comment}
The Liouville equation $i\hbar\dot\rho=[h,\rho]$ yields
\begin{equation}
\begin{aligned}
\rho^{(1),E}_{nm}(t)
&=-\frac{ef_{nm}r^b_{nm}}{\hbar}
\sum_{\alpha=\pm1}
\frac{E^b(\alpha\omega)e^{-i\alpha\omega t}}
{\omega_{nm}-\alpha\omega-i\eta_E},\\
\rho^{(1),B}_{nm}(t)
&=-\frac{\lambda_Zf_{nm}\sigma^\beta_{nm}}{\hbar}
\sum_{\alpha=\pm1}
\frac{B^\beta(\alpha\omega)e^{-i\alpha\omega t}}
{\omega_{nm}-\alpha\omega-i\eta_B}.
\end{aligned}
\label{eq:rho1}
\end{equation}

Here $\mathcal E^b=E^b(+\omega)$ denotes the positive-frequency
field amplitude, with $E^b(-\omega)=\mathcal E^{b*}$ for real fields;
the same convention applies to $B^\beta$.
Substituting these first-order responses into the Liouville
equation gives, at mixed second order,
\end{comment}
%\begin{equation}
%\begin{aligned}
%(\partial_t+i\omega_{nm})\rho^{(2),EB}_{nm} ={}&-\frac{eE^b}{\hbar}(\rho^{(1),B^\beta}_{nm})_{;b}\\ &+\frac{ieE^b}{\hbar}[r^b,\rho^{(1),B^\beta}]_{nm}\\ &+\frac{i\lambda_ZB^\beta}{\hbar} [\sigma^\beta,\rho^{(1),E^b}]_{nm}.
%\end{aligned}
%\label{eqS:rho2-eom}
%\end{equation}
\begin{equation}
\begin{aligned}
&(\partial_t+i\omega_{nm})\rho^{(2),EB}_{nm}
=-\frac{eE^b}{\hbar}(\rho^{(1),B^\beta}_{nm})_{;b}\\
&\quad+\frac{i}{\hbar}\Big(
eE^b[r^b,\rho^{(1),B^\beta}]_{nm}
+\lambda_ZB^\beta[\sigma^\beta,\rho^{(1),E^b}]_{nm}
\Big).
\end{aligned}
\label{eq:rho2-eom}
\end{equation}

For $\eta_E=\eta_B=\eta>0$, Eq.~\eqref{eq:rho2-eom}
gives the dc response
\begin{equation}
\rho^{(2),EB}_{nm,\mathrm{dc}}(t)
=\sum_{b\beta}
\bigl[\mathcal E^{b*}\mathcal B^\beta K_{nm}
+\mathcal E^b\mathcal B^{\beta*}K_{mn}^*\bigr],
\label{eq:rho-dc-pairings}
\end{equation}
where $D_{nm}^{\pm}=\omega_{nm}\pm\omega-i\eta$ and
%Solving Eq.~\eqref{eq:rho2-eom} with $\eta_E=\eta_B=\eta>0$ gives the dc response
\begin{equation}
\begin{aligned}
K_{nm}=-\frac{e\lambda_Z}{\hbar^2(\omega_{nm}-2i\eta)}
\Bigg\{&i\left[\frac{f_{nm}\sigma^\beta_{nm}}{D_{nm}^{-}}\right]_{;b}\\
{}+\sum_l\Big[\frac{r^b_{nl}f_{lm}\sigma^\beta_{lm}}{D_{lm}^{-}}
&-\frac{f_{nl}\sigma^\beta_{nl}r^b_{lm}}{D_{nl}^{-}}\Big]\\
{}+\sum_l\Big[\frac{\sigma^\beta_{nl}f_{lm}r^b_{lm}}{D_{lm}^{+}}
&-\frac{f_{nl}r^b_{nl}\sigma^\beta_{lm}}{D_{nl}^{+}}\Big]\Bigg\}.
\end{aligned}
\label{eq:K-selected}
\end{equation}

Writing the current-flow index as $a$, the first index of the subscript, tracing the dc density matrix gives
	\begin{equation}
		\begin{aligned}
			\mathcal C^s_{ab{\beta}}
			&=\frac1V\sum_{nm}\int dk\,(j_a^s)_{mn}K_{nm},\\
			J_{a,\mathrm{dc}}^{s,EB}
			&=2\operatorname{Re}\sum_{b\beta}
			\mathcal C^s_{ab{\beta}}\mathcal E^{b*}\mathcal B^\beta.
		\end{aligned}
		\label{eq:dc-current-trace}
	\end{equation}
	Here $s$ denotes spin polarization and the $\beta$ identifies the magnetic-field index. To extract the resonant injection and shift spin currents, we separate $\mathcal C=\mathcal C_{\mathrm d}+\mathcal C_{\mathrm o}$ into population ($n=m$) and coherence ($n\ne m$) contributions. 
	
\textit{Injection spin current}---
For the population contribution ($n=m$), the derivative term
vanishes and the output denominator reduces to $-2i\eta$.
Combining the input poles with
$(x+i\eta)^{-1}-(x-i\eta)^{-1}=-2i\eta/(x^2+\eta^2)$,
then exchanging $n\leftrightarrow m$ in one resonance orientation, gives
\begin{equation}
\begin{aligned}
\mathcal C_{\mathrm d}
=-\frac{e\lambda_Z}{\hbar^2V}\sum_{n\ne m}\int dk\,
&\frac{f_{nm}r^b_{nm}\sigma^\beta_{mn}}
{(\omega_{nm}+\omega)^2+\eta^2}\\
&\times[(j_a^s)_{nn}-(j_a^s)_{mm}].
\end{aligned}
\label{eqS:diag-trace}
\end{equation}

The current difference reflects opposite population changes in the two bands. Retaining the resonant contribution and setting $\eta=1/(2\tau)$, where $\tau$ is the population relaxation time, we obtain the steady-state injection coefficient.
\begin{equation}
\begin{aligned}
\mathcal I_{ab {\beta}}^{s}(\omega)
={}&-\frac{2\pi e\lambda_Z\tau}{\hbar^2V}
\sum_{n\ne m}\int dk\,f_{nm}r^b_{nm}\sigma^\beta_{mn}\\
&\times[(j_a^s)_{nn}-(j_a^s)_{mm}]
\delta(\omega_{nm}+\omega).
\end{aligned}
\label{eq:Icompact}
\end{equation}
where $(j_a^s)_{nn}$ follows from Eq.~\eqref{eq:spin_matrix}. A steady-state injection current therefore requires
population relaxation.

\textit{Shift spin current}---
We focus on the resonant coherence response ($n\ne m$), retaining the $\delta$ and $\delta'$ terms in the input-pole identities $(x-i0^+)^{-1}=\mathcal P(1/x)+i\pi\delta(x)$ and $(x-i0^+)^{-2}=\operatorname{Pf}(1/x^2)-i\pi\delta'(x)$.
The principal-value and finite-part terms belong to the dispersive remainder. For nonzero output gaps, $(\omega_{nm}-2i\eta)^{-1}\to\omega_{nm}^{-1}$ as $\eta\to0^+$, and we introduce 
\begin{equation}
Q_{a,mn}^s=-\frac{(j_a^s)_{mn}}{\omega_{mn}}\quad(m\ne n),
\label{eq:Q}
\end{equation}
for $n\ne m$, with $Q^s_{a,nn}=0$. The finite resonant response then takes the form $\mathcal S=\mathcal S_1+\mathcal S_2+\mathcal S_3$, where the three contributions correspond to the three sources in Eq.~\eqref{eq:K-selected}, respectively.
\begin{equation}
\begin{aligned}
\mathcal S_1
&=\frac{\pi e\lambda_Z}{\hbar^2V}
\sum_{nm}\int dk\,Q^s_{a,mn}
[f_{nm}\sigma^\beta_{nm}\delta(\omega_{nm}-\omega)]_{;b},
\\[2pt]
\mathcal S_2
&=-\frac{i\pi e\lambda_Z}{\hbar^2V}
\sum_{nml}\int dk\,Q^s_{a,mn}\\
&\times
[r^b_{nl}f_{lm}\sigma^\beta_{lm}\delta(\omega_{lm}-\omega)
-f_{nl}\sigma^\beta_{nl}r^b_{lm}\delta(\omega_{nl}-\omega)],
\\[2pt]
\mathcal S_3
&=-\frac{i\pi e\lambda_Z}{\hbar^2V}
\sum_{nml}\int dk\,Q^s_{a,mn}\\
&\times
[\sigma^\beta_{nl}f_{lm}r^b_{lm}\delta(\omega_{lm}+\omega)
-f_{nl}r^b_{nl}\sigma^\beta_{lm}\delta(\omega_{nl}+\omega)].
\end{aligned}
\label{eqS:shift-sources}
\end{equation}
We first simplify $\mathcal S_1$. Its $\delta$ and $\delta'$
terms combine into a covariant derivative.
Writing
$Y_{nm}=f_{nm}\sigma^\beta_{nm}\delta(\omega_{nm}-\omega)$,
covariant integration by parts over the Brillouin zone gives
\begin{equation}
\int dk\,Q^s_{a,mn}Y_{nm;b}
=-\int dk\,(Q^s_{a;b})_{mn}Y_{nm}.
\label{eqS:covariant-ibp}
\end{equation}
The Berry connections cancel in the product, so this step
retains the $\delta'$ contribution. Relabeling the intermediate
bands in $\mathcal S_2$ and $\mathcal S_3$ then yields
\begin{equation}
\begin{aligned}
\mathcal S_1+\mathcal S_2
={}&-\frac{\pi e\lambda_Z}{\hbar^2V}
\sum_{nm}\int dk\,f_{nm}\sigma^\beta_{nm}\\
&\times\delta(\omega_{nm}-\omega)
\big[(Q^s_{a;b})_{mn}+i[Q^s_a,r^b]_{mn}\big],\\
\mathcal S_3
={}&-\frac{i\pi e\lambda_Z}{\hbar^2V}
\sum_{nm}\int dk\,f_{nm}r^b_{nm}\\
&\times[Q^s_a,\sigma^\beta]_{mn}
\delta(\omega_{nm}+\omega).
\end{aligned}
\label{eqS:shift-contracted}
\end{equation}
Exchanging $n\leftrightarrow m$ in the first expression
places all terms on the same resonance, giving
\begin{equation}
\begin{aligned}
\mathcal S_{ab{\beta}}^{s}(\omega)
={}&\frac{\pi e\lambda_Z}{\hbar^2V}
\sum_{n\ne m}\int dk\,f_{nm}\delta(\omega_{nm}+\omega)\\
&\times\Big\{\sigma^\beta_{mn}
[(Q^s_{a;b})_{nm}+i[Q^s_a,r^b]_{nm}]\\
&\hspace{24pt}-ir^b_{nm}[Q^s_a,\sigma^\beta]_{mn}\Big\}.
\end{aligned}
\label{eq:Scompact}
\end{equation}
Here $[Q,O]_{mn}=\sum_l(Q_{ml}O_{ln}-O_{ml}Q_{ln})$, with $Q_{ml}$ defined in Eq.~\eqref{eq:Q}, The covariant derivative of $Q$ reads
\begin{equation}
(Q^s_{a;b})_{nm}=-\frac{(j_a^s)_{nm;b}}{\omega_{nm}}
+\frac{(j_a^s)_{nm}(v_n^b-v_m^b)}{\omega_{nm}^2}.
\label{eqS:Q-derivative}
\end{equation}
where $(j_a^s)_{mn}$ is given by Eq.~\eqref{eq:spin_matrix}.
The resulting shift spin current is
$J_{a,\mathrm{shift}}^{s,EB}
=2\operatorname{Re}\sum_{b\beta}
\mathcal S_{ab\beta}^{s}\mathcal E^{b*}\mathcal B^\beta$.

Eqs.~\eqref{eq:Icompact} and~\eqref{eq:Scompact} give the injection and shift spin-current conductivities, respectively, and constitute the central results of this work. Detailed derivations are provided in the Supplemental Material~\cite{supp}. As expected, the shift spin current is independent of relaxation time.

\textcolor{blue}{\textbf{Symmetry constraints}}---We classify the electric-dipole ($EE$) and mixed electric--Zeeman ($EB$) spin responses under inversion $\mathcal P$, time reversal $\mathcal T$, and their combinations $\mathcal {PT}$. The spin-current operator $j_a^s=\{j_a,\sigma^s\}/2$ is odd under $\mathcal P$ and even under $\mathcal T$. For $\mathcal P$, since $EE$ is inversion even whereas $EB$ is odd, inversion symmetry forbids conventional spin shift and injection currents but permits both polarization channels of their Zeeman counterparts.

To establish the antiunitary selection rules, we introduce the
transition kernels $\mathcal R_{\mathrm S}^{XY}$ and
$\mathcal R_{\mathrm I}^{XY}$, with $XY=EE,EB$.
The $EB$ kernels in  Eqs.~\eqref{eq:Scompact}
give $\mathcal R_{\mathrm S,nm}^{EB}
=\sigma^\beta_{mn}((Q^s_{a;b})_{nm}+i[Q^s_a,r^b]_{nm})
-i r^b_{nm}[Q^s_a,\sigma^\beta]_{mn}$, and $\mathcal R_{\mathrm I,nm}^{EB}=\Delta j_{a,nm}^s\,r^b_{nm}\sigma^\beta_{mn}$. For $EE$ excitation, the resonant expressions in~\cite{Lihm2022}, with $\mathcal R_{\mathrm I,nm}^{EE}=
\Delta j_{a,nm}^s\,r^b_{mn}r^c_{nm}$ and $\mathcal R_{\mathrm S,nm}^{EE}
=\frac{-i}{\omega_{nm}^2}(v^b_{mn}d_{nm}^{s;c;a}-v^c_{nm}d_{mn}^{s;b;a})$.

An antiunitary symmetry $X$ maps $|n,\mathbf k\rangle$ to
an equal-energy partner $|n_X,\mathbf k_X\rangle=X|n,\mathbf k\rangle$,
with $\mathbf k_X=-\mathbf k$ for $\mathcal T$ and
$\mathbf k_X=\mathbf k$ for $\mathcal{PT}$.
Partner matrix elements obey
\begin{equation}
\langle Xn|\mathcal O|Xm\rangle
=\langle n|X^{-1}\mathcal OX|m\rangle^*.
\label{eq:spin_antiunitary_short}
\end{equation}
The transitions $(n,m,\mathbf k)$ and $(n_X,m_X,\mathbf k_X)$
have identical occupation and resonance weights.
Their pair average therefore selects $\operatorname{Re}\mathcal R$
when $\mathcal R_X=\mathcal R^*$ and
$i\operatorname{Im}\mathcal R$ when
$\mathcal R_X=-\mathcal R^*$, defining the L and C channels,
respectively.

Under $\mathcal T$, the partner kernels satisfy
\begin{equation}
\begin{aligned}
(\mathcal R_{\mathrm S}^{EE},\mathcal R_{\mathrm I}^{EE})_{\mathcal T}
&=\bigl(-(\mathcal R_{\mathrm S}^{EE})^*,
         (\mathcal R_{\mathrm I}^{EE})^*\bigr),\\
(\mathcal R_{\mathrm S}^{EB},\mathcal R_{\mathrm I}^{EB})_{\mathcal T}
&=\bigl((\mathcal R_{\mathrm S}^{EB})^*,
        -(\mathcal R_{\mathrm I}^{EB})^*\bigr).
\end{aligned}
\label{eq:spin_T_kernels_short}
\end{equation}
Thus, the Brillouin-zone integration retains only the imaginary and  real parts of the $EE$ spin-shift and spin-injection respectively.
For $EB$, only the real part contributes to spin shift and the imaginary part to spin injection.
Under $\mathcal{PT}$, both $EE$ and $EB$ retain the real part for spin shift and the imaginary part for spin injection.
Table~\ref{tab:spin_selection_short} summarizes the selection rules, with the $L$ and $C$ channels interchanged relative to those for Zeeman charge shift and injection currents~\cite{Dong2026zeeman,Xiang2026}. Detailed definitions and symmetry transformations are provided in the Supplemental Material~\cite{supp}.

\begin{table}[t]
  \centering
  \caption{Channels allowed by each symmetry separately. L and C denote symmetry-allowed response for linearly and circularly polarized light, respectively.}
  \label{tab:spin_selection_short}
  \setlength{\tabcolsep}{3pt}
  \renewcommand{\arraystretch}{1.0}
  \setlength{\aboverulesep}{0pt}
  \setlength{\belowrulesep}{0pt}
  \small
  \begin{tabular}{@{}c cc cc@{}}
    \toprule
    & \multicolumn{2}{c}{$EE$}
    & \multicolumn{2}{c}{$EB$} \\
    \cmidrule(lr){2-3}\cmidrule(lr){4-5}
    Symmetry &spin shift &spin injection &spin shift &spin injection \\
    \hline
    $\mathcal P$ & -- & -- & L/C & L/C \\
    $\mathcal T$ & C & L & L & C \\
    $\mathcal{PT}$ & L & C & L & C \\
    \hline\hline
  \end{tabular}
\end{table}

\begin{figure}[!b]
%\hspace*{-3cm} 
%\vspace{-1cm}
\includegraphics[width=8.5 cm]{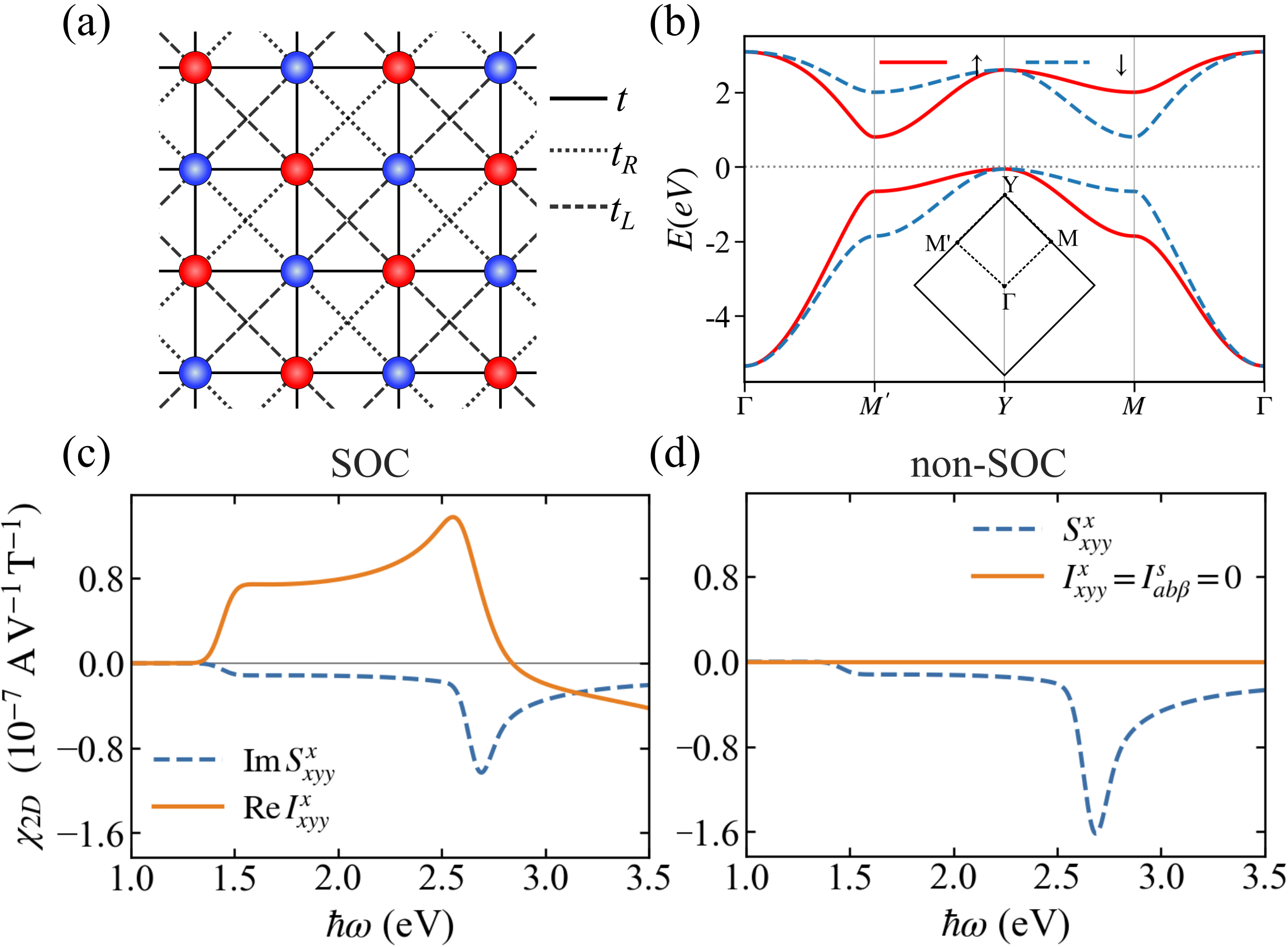}
\caption{
Zeeman spin photocurrents in a two-dimensional altermagnet.
(a) Lattice with antiparallel sublattice moments along $z$ and hopping amplitudes $t$, $t_R$, and $t_L$.
(b) Spin-resolved bands without SOC. Inset: Brillouin zone and high-symmetry path.
(c,d) Share the same $y$-axis scale. Zeeman shift (blue) and steady-state injection (red, $\tau=0.1\,\mathrm{ps}$) spin-current conductivities with Rashba SOC ($\lambda_R=0.08\,\mathrm{eV}$) and without SOC, respectively.}
\label{Fig1}
\end{figure}

\textcolor{blue}{\textbf{Zeeman spin photocurrents in a centrosymmetric altermagnetic model}}---
\begin{comment}
To illustrate the preceding theory, we consider a two-dimensional
square-lattice altermagnet with two sublattices carrying opposite
magnetic moments along $z$, as shown in
Fig.~\ref{fig:AM_spin_response}(a).
The nearest-neighbor hopping is $t$, while the diagonal hoppings
$t_L$ and $t_R$ interchange between the two sublattices.
The Hamiltonian of the system is

The symmetry analysis above shows that Zeeman spin photocurrents are allowed in centrosymmetric systems, irrespective of spin splitting. Since many altermagnets are centrosymmetric, we illustrate the theory using a simple two-dimensional square-lattice altermagnetic model with two sublattices carrying opposite magnetic moments along $z$, as shown in Fig.~\ref{Fig1}(a). 
The Hamiltonian reads
\end{comment}
The symmetry analysis above shows that Zeeman spin photocurrents are allowed in centrosymmetric systems, irrespective of spin splitting. Since many altermagnets are centrosymmetric, we consider a %simple
two-dimensional square-lattice altermagnetic model with two sublattices carrying opposite moments along $z$, as shown in Fig.~\ref{Fig1}(a). The Hamiltonian reads
\begin{equation}
\hat H=-\sum_{i,j,s}t_{ij}
(c^\dagger_{is}c_{js}+\mathrm{H.c.})
+U\sum_i n_{i\uparrow}n_{i\downarrow},
\label{eq:AM-hamiltonian}
\end{equation}
where $U$ is the on-site repulsion, $t_{ij}=t$ for nearest neighbors, and the diagonal hoppings $t_L$ and $t_R$ are interchanged between % the two 
sublattices. Following Ref.~\cite{Das2024}, we treat the interaction within the %a collinear 
Hartree--Fock approximation with the altermagnetic order parameter $m$.

\begin{comment}
The nearest-neighbor hopping is $t$, while the diagonal hoppings $t_L$ and $t_R$ are interchanged between the two sublattices. The Hamiltonian reads
\begin{align}
        \hat{H} = -\sum_{i,j,s} t_{ij} (c^\dagger_{is} c_{js} + h.c.) + U \sum_i n_{i\uparrow} n_{i\downarrow}.
        \label{eq:AM-hamiltonian}
\end{align}
Where $U$ is the on-site repulsion.
Following ~\cite{Das2024}, we treat the interaction within
a collinear Hartree--Fock approximation by introducing an
altermagnetic order parameter $m$.
This yields opposite exchange fields of magnitude $J=Um$
on the two sublattices.

In the absence of SOC, spin and spatial transformations are
independent.
In the present coordinate convention, the spin
point group is ${}^{2}4/{}^{1}m{}^{2}m{}^{1}m$, where superscripts
$1$ and $2$ denote the spin identity and a $\pi$ spin rotation
about an axis perpendicular to the N\'eel vector, respectively.
Spatial inversion is preserved, and $\sigma_z$ remains conserved.
Figure~\ref{Fig1}(b) shows the SOC-free bands,
whose alternating spin splitting reflects the combined
spin-flip and spatial-rotation symmetry.
\end{comment}

Without SOC, spin and spatial transformations are independent. In our coordinate convention, the spin point group is ${}^{2}4/{}^{1}m{}^{2}m{}^{1}m$, where superscripts $1$ and $2$ denote the spin identity and a $\pi$ spin rotation about an axis perpendicular to the N\'eel vector, respectively. The model preserves inversion and conserves $\sigma_z$. The bands in Fig.~\ref{Fig1}(b) exhibit alternating spin splitting, reflecting the combined spin-flip and spatial-rotation symmetry.

At first glance, Eqs.~\eqref{eq:Icompact} and~\eqref{eq:Scompact} appear to require spin mixing for nonzero $\sigma_{mn}$. Motivated by this expectation, we introduce the Rashba SOC term
%Eqs.~\eqref{eq:Icompact} and~\eqref{eq:Scompact} depend explicitly on the spin matrix elements $\sigma_{mn}$. We therefore introduce Rashba SOC to examine the role of spin mixing.
\begin{equation}
h_R(\mathbf k)
=\lambda_R\tau_x
\left[\sin(k_xa)\sigma_y-\sin(k_ya)\sigma_x\right]
\label{eq:AM-Rashba}
\end{equation}
to the mean-field Hamiltonian~\eqref{eq:AM-hamiltonian}.
Here $\tau_x$ acts in sublattice space and $a$ is the lattice
spacing.
With the N'eel vector fixed along $z$, Rashba SOC reduces the symmetry to the magnetic point group $4'mm'$.
%This term breaks inversion and the independence of spin-rotation and real space rotation.
%For the N\'eel vector fixed in $z$-direction, the remaining magnetic point group is $4'mm'$. We compare $\lambda_R=0$ and $0.08\,\mathrm{eV}$ while keeping other parameters fixed.

With Rashba SOC, both shift and injection responses are finite, as shown in Fig.~\ref{Fig1}(c). Unexpectedly, the two independent shift spectra, $\operatorname{Re}S^x_{xxx}$ and $\operatorname{Im}S^x_{xyy}$, remain finite even without SOC, whereas all injection components vanish, as shown in Fig.~\ref{Fig1}(d). The surviving shift current carries spin polarization transverse to the N\'eel vector. 
\begin{comment}
With Rashba SOC, both shift and injection responses are finite, as shown in Fig.~\ref{Fig1}(c). Unexpectedly, the two independent shift spectra, $\operatorname{Re}S^x_{xxx}$ and $\operatorname{Im}S^x_{xyy}$, remain finite even without SOC, whereas all injection components vanish, as shown in Fig.~\ref{Fig1}(d). The surviving shift current carries spin polarization transverse to the N\'eel vector.
\end{comment}

Equilibrium spin mixing is therefore required for injection in this model, but not for the transverse shift response. The latter originates from optically induced interband spin coherence. We next examine the selection rules governing this SOC-free response.

\begin{comment}
The corresponding mixed $EB$ spin current responses are calculated using
Eqs.~\eqref{eq:Icompact} and~\eqref{eq:Scompact}.
For the figure, we write
$S^s_{ab\beta}\equiv\mathcal S_a^{s;b\beta}$ and
$I^s_{ab\beta}\equiv\mathcal I_a^{s;b\beta}(\omega;\tau)$,
with the relaxation time $\tau=0.1\,\mathrm{ps}$ already
included in $I$.

Without SOC, the two independent shift spectra,
$\operatorname{Re}S^x_{xxx}$ and
$\operatorname{Im}S^x_{xyy}$, remain finite, whereas all
injection components vanish
[Fig.~\ref{fig:AM_spin_response}(c)].
The surviving shift response carries spin polarization
transverse to the N\'eel vector.
Rashba SOC modifies these spectra and enables finite
injection responses
[Fig.~\ref{fig:AM_spin_response}(d)].
Thus, spin mixing in the equilibrium bands is required for
injection in this model, but not for the transverse spin-shift
response.
The latter originates from optically induced interband spin
coherence, whose SOC-free selection rules we examine next.
\end{comment}

\textcolor{blue}{\textbf{Zeeman-shift-induced pure spin currents without SOC}}---
Counterintuitively, pure Zeeman shift spin currents can arise in collinear magnets without SOC or equilibrium spin mixing. This mechanism offers a route to optical spin-current sources based on both conventional antiferromagnets with spin-degenerate bands and altermagnets.

Choosing $z$ along the N\'eel vector, we write the unperturbed Hamiltonian as $H_0(\mathbf k)=h(\mathbf k)\otimes I+M(\mathbf k)\otimes\sigma^z$.
%With the N\'eel vector fixed along $z$, the unperturbed Hamiltonian reads
%\begin{equation}
%H_0(\mathbf k)=h(\mathbf k)\otimes I+M(\mathbf k)\otimes\sigma^z.
%\label{eq:nonrelativistic-hamiltonian}
%\end{equation}
%The Hamiltonian separates into two independent spin blocks $H_{\uparrow,\downarrow}(\mathbf k)=h(\mathbf k)\pm M(\mathbf k)$, so the spin projection along $z$ remains a good quantum number. Hereafter, $n,m,l$ denote orbital-band indices and $s=\uparrow,\downarrow$ labels spin, so that $(n,s)$ specifies a complete eigenstate. The eigenstates factorize as $|u_{ns}\rangle=|\psi_n^{(s)}\rangle\otimes|\chi_s\rangle$.
%Consequently, the electric vertex preserves spin, whereas the transverse Zeeman vertices connect opposite spins through orbital overlaps $\langle\psi_n^{(\uparrow)}|\psi_m^{(\downarrow)}\rangle$, which need not vanish. The longitudinal vertex $\sigma^z$ is diagonal in the complete  band--spin basis.
The Hamiltonian separates into spin blocks
$H_{\uparrow,\downarrow}(\mathbf k)=h(\mathbf k)\pm M(\mathbf k)$,
with eigenstates
$|u_{ns}\rangle=|\psi_n^{(s)}\rangle\otimes|\chi_s\rangle$.
Here, $n,m,l$ label bands within each block, and $s=\uparrow,\downarrow$ denotes the conserved spin projection along $z$.
The electric vertex $r^b$ preserves spin, whereas $\sigma^{x,y}$ connect opposite spins, like $\langle u_{n\uparrow}|\sigma^x|u_{m\downarrow}\rangle = \langle\psi_n^{(\uparrow)}|\psi_m^{(\downarrow)}\rangle\langle \uparrow|\sigma^x|\downarrow\rangle$, while $\langle \uparrow|\sigma^x|\downarrow\rangle = 1$ and
$\langle\psi_n^{(\uparrow)}|\psi_m^{(\downarrow)}\rangle$ need not vanish between different spin blocks.
The longitudinal vertex $\sigma^z$ is diagonal in both band and spin indices.

\begin{figure}[!t]
\includegraphics[width=7cm]{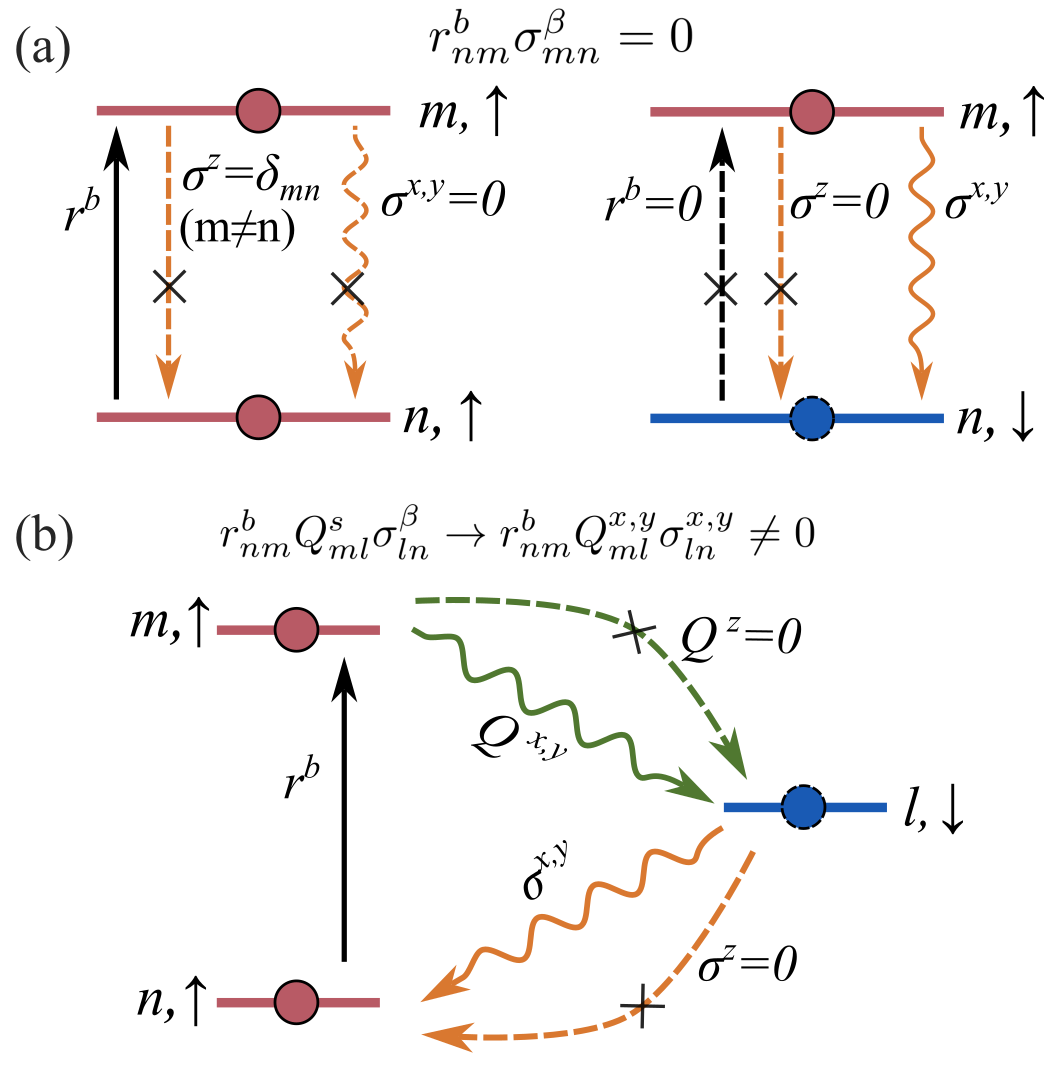}
%\vspace{-1.2cm}
\caption{
Spin-selection rules for SOC-free Zeeman spin photocurrents.
Red and blue levels represent opposite eigenstates of $\sigma^z$.
(a) The interband product $r^b_{nm}\sigma^\beta_{mn}$ vanishes, forbidding spin injection.
(b) Two spin flips at the transverse Zeeman and spin-current vertices enable a finite shift spin current.
Black, orange, and green arrows denote the $r$, $\sigma$, and $Q$ vertices, respectively. Solid and dashed arrows indicate allowed and forbidden matrix elements.
}
\label{Fig2}
\end{figure}

The injection response involves the product
$r^b_{ns,ms'}\sigma^\beta_{ms',ns}$.
The electric matrix element vanishes between opposite-spin states, whereas the interband Zeeman matrix element vanishes between same-spin states. Consequently, both charge and spin injection vanish [Fig.~\ref{Fig2}(a)].

The spin-shift current components with $\alpha=z$ or $\beta=z$ vanish by spin selection.
For $\beta=z$, $\sigma^z$ is diagonal, while $[Q_a^\alpha,\sigma^z]$ vanishes between the same-spin states connected by $r^b$.
For $\alpha=z$ and $\beta=x,y$, spin-current vertices preserve spin, leaving an uncompensated Zeeman-induced spin flip.
Thus, $\mathcal S_{abz}^{\alpha}=\mathcal S_{ab\beta}^{z}=0,\
(\alpha,\beta=x,y,z).$

For transverse $\alpha,\beta=x,y$, both $Q_a^\alpha$ and $\sigma^\beta$ flip spin. Their commutator can therefore connect same-spin states through two successive spin flips and combine with the electric vertex [Fig.~\ref{Fig2}(b)].
The vertices $Q_{a;b}^\alpha$ and $[Q_a^\alpha,r^b]$ also flip spin and can pair with $\sigma^\beta$.
Thus, both parts of the transverse spin-shift kernel can remain finite despite the vanishing injection product.

\begin{table}[!t]
  \centering
  \small
  \begingroup
  \setlength{\tabcolsep}{1.5pt}
  \newcommand{\ZSmatrix}[1]{%
    \begingroup
    \setlength{\arraycolsep}{1.5pt}%
    \renewcommand{\arraystretch}{0.8}%
    \begin{pmatrix}#1\end{pmatrix}%
    \endgroup}
\caption{
%Symmetry-allowed Zeeman spin-shift response in centrosymmetric spin point groups. L and C give the real $\chi^x_{abx}$ and purely imaginary $\chi^x_{aby}$ matrices, respectively. Distinct symbols in each row are independent.
Symmetry-allowed Zeeman spin-shift response in centrosymmetric spin point-group. The L and C columns give the real $(S^x_{abx})$ and purely imaginary $(S^x_{aby})$ matrices,     respectively.     Distinct component symbols within each row are independent.
}
  \label{tab:centrosymmetric-zeeman-spin-shift}
  \begin{tabular}{@{}ccc@{}}
    \toprule
    Spin point group & L channel $(\alpha=\beta)$ & C channel $(\alpha\ne\beta)$ \\
    \midrule
    ${}^{2}2/{}^{2}m$
    & $\ZSmatrix{S^x_{xxx}&S^x_{xyx}&0\\S^x_{yxx}&S^x_{yyx}&0\\0&0&S^x_{zzx}}$
    & $\ZSmatrix{0&0&S^x_{xzy}\\0&0&S^x_{yzy}\\S^x_{zxy}&S^x_{zyy}&0}$ \\
    \addlinespace[2pt]
    ${}^{2}m{}^{2}m{}^{1}m$
    & $\ZSmatrix{S^x_{xxx}&0&0\\0&S^x_{yyx}&0\\0&0&S^x_{zzx}}$
    & $\ZSmatrix{0&S^x_{xyy}&0\\S^x_{yxy}&0&0\\0&0&0}$ \\
    \addlinespace[2pt]
    ${}^{2}4/{}^{1}m$
    & $\ZSmatrix{S^x_{xxx}&S^x_{xyx}&0\\-S^x_{xyx}&S^x_{xxx}&0\\0&0&S^x_{zzx}}$
    & $\ZSmatrix{S^x_{xxy}&S^x_{xyy}&0\\S^x_{xyy}&-S^x_{xxy}&0\\0&0&0}$ \\
    \addlinespace[2pt]
    ${}^{1}4/{}^{1}m{}^{2}m{}^{2}m$
    & $\ZSmatrix{S^x_{xxx}&0&0\\0&S^x_{xxx}&0\\0&0&S^x_{zzx}}$
    & $\ZSmatrix{0&S^x_{xyy}&0\\-S^x_{xyy}&0&0\\0&0&0}$ \\
    \addlinespace[2pt]
    ${}^{2}4/{}^{1}m{}^{2}m{}^{1}m$
    & $\ZSmatrix{S^x_{xxx}&0&0\\0&S^x_{xxx}&0\\0&0&S^x_{zzx}}$
    & $\ZSmatrix{0&S^x_{xyy}&0\\S^x_{xyy}&0&0\\0&0&0}$ \\
    \addlinespace[2pt]
    ${}^{1}\bar{3}{}^{2}m$
    & $\ZSmatrix{S^x_{xxx}&0&0\\0&S^x_{xxx}&0\\0&0&S^x_{zzx}}$
    & $\ZSmatrix{0&S^x_{xyy}&0\\-S^x_{xyy}&0&0\\0&0&0}$ \\
    \addlinespace[2pt]
    ${}^{2}6/{}^{2}m$
    & $\ZSmatrix{S^x_{xxx}&S^x_{xyx}&0\\-S^x_{xyx}&S^x_{xxx}&0\\0&0&S^x_{zzx}}$
    & $\mathbf /$ \\
    \addlinespace[2pt]
    ${}^{1}6/{}^{1}m{}^{2}m{}^{2}m$
    & $\ZSmatrix{S^x_{xxx}&0&0\\0&S^x_{xxx}&0\\0&0&S^x_{zzx}}$
    & $\ZSmatrix{0&S^x_{xyy}&0\\-S^x_{xyy}&0&0\\0&0&0}$ \\
    \addlinespace[2pt]
    ${}^{2}6/{}^{2}m{}^{2}m{}^{1}m$
    & $\ZSmatrix{S^x_{xxx}&0&0\\0&S^x_{xxx}&0\\0&0&S^x_{zzx}}$
    & $\mathbf /$ \\
    \addlinespace[2pt]
    ${}^{1}m{}^{1}\bar{3}{}^{2}m$
    & $\ZSmatrix{S^x_{xxx}&0&0\\0&S^x_{xxx}&0\\0&0&S^x_{xxx}}$
    & $\mathbf /$ \\
    \bottomrule
  \end{tabular}
  \endgroup
\end{table}

Spin conservation further constrains the transverse spin-shift response. For $X=\sigma,Q_a$, the spin-flip blocks obey
$X^y_{\uparrow\downarrow}=-iX^x_{\uparrow\downarrow}$ and
$X^y_{\downarrow\uparrow}=iX^x_{\downarrow\uparrow}$.
Applying these relations to both terms of the shift kernel yields
\begin{equation}
\left(S^\alpha_{ab\beta}\right)_{\alpha,\beta=x,y}
=
\begin{pmatrix}
S^x_{abx}&S^x_{aby}\\
-S^x_{aby}&S^x_{abx}
\end{pmatrix}.
\label{eq:collinear-transverse-spin-structure}
\end{equation}
Together with the antiunitary symmetry $[C_2\Vert T]$, where $C_2$ is a $\pi$ spin rotation about an axis perpendicular to the N\'eel vector, this structure requires
$S^\alpha_{ab\beta}=(2\delta_{\alpha\beta}-1)(S^\alpha_{ab\beta})^*$
for $\alpha,\beta=x,y$.
The $\alpha=\beta$ components are therefore real and couple to the real part of $\mathcal E^{b*}\mathcal B^\beta$ (L channel), whereas the $\alpha\ne\beta$ components are purely imaginary and couple to its imaginary part (C channel).

Table~\ref{tab:centrosymmetric-zeeman-spin-shift} summarizes the symmetry-allowed Zeeman spin-shift currents in centrosymmetric spin point groups. The L channel is allowed in every listed group, whereas the C channel is forbidden in several groups. In particular, the cubic group ${}^{1}m{}^{1}\bar{3}{}^{2}m$ permits only a spatially isotropic L-channel response with a single independent coefficient.

\begin{comment}
Spin conservation also relates the allowed transverse components.
The spin-flip blocks satisfy
$X^y_{\uparrow\downarrow}=-iX^x_{\uparrow\downarrow}$ and
$X^y_{\downarrow\uparrow}=iX^x_{\downarrow\uparrow}$
for $X=\sigma,Q_a$.
Substitution into both terms of the shift kernel therefore gives
\begin{equation}
  \left(\chi^\alpha_{ab\beta}\right)_{\alpha,\beta=x,y}
  =\begin{pmatrix}
    \chi^x_{abx}&\chi^x_{aby}\\
    -\chi^x_{aby}&\chi^x_{abx}
  \end{pmatrix}.
  \label{eq:collinear-transverse-spin-structure}
\end{equation}
The antiunitary symmetry $[C_2\Vert T]$, where $C_2$ is a
$\pi$ spin rotation about any axis perpendicular to the N\'eel vector,
relates transition matrix elements to their complex conjugates.
Combined with the transverse spin structure, it imposes
$\chi^\alpha_{ab\beta}=(2\delta_{\alpha\beta}-1)
(\chi^\alpha_{ab\beta})^*$ for $\alpha,\beta=x,y$.
The $\alpha=\beta$ and $\alpha\ne\beta$ components are therefore
real and purely imaginary, respectively, coupling to the real
(L channel) and imaginary (C channel) parts of
$\mathcal E^{b*}\mathcal B^\beta$.     
\end{comment}

\begin{figure}[!t]
\includegraphics[width=8.5cm]{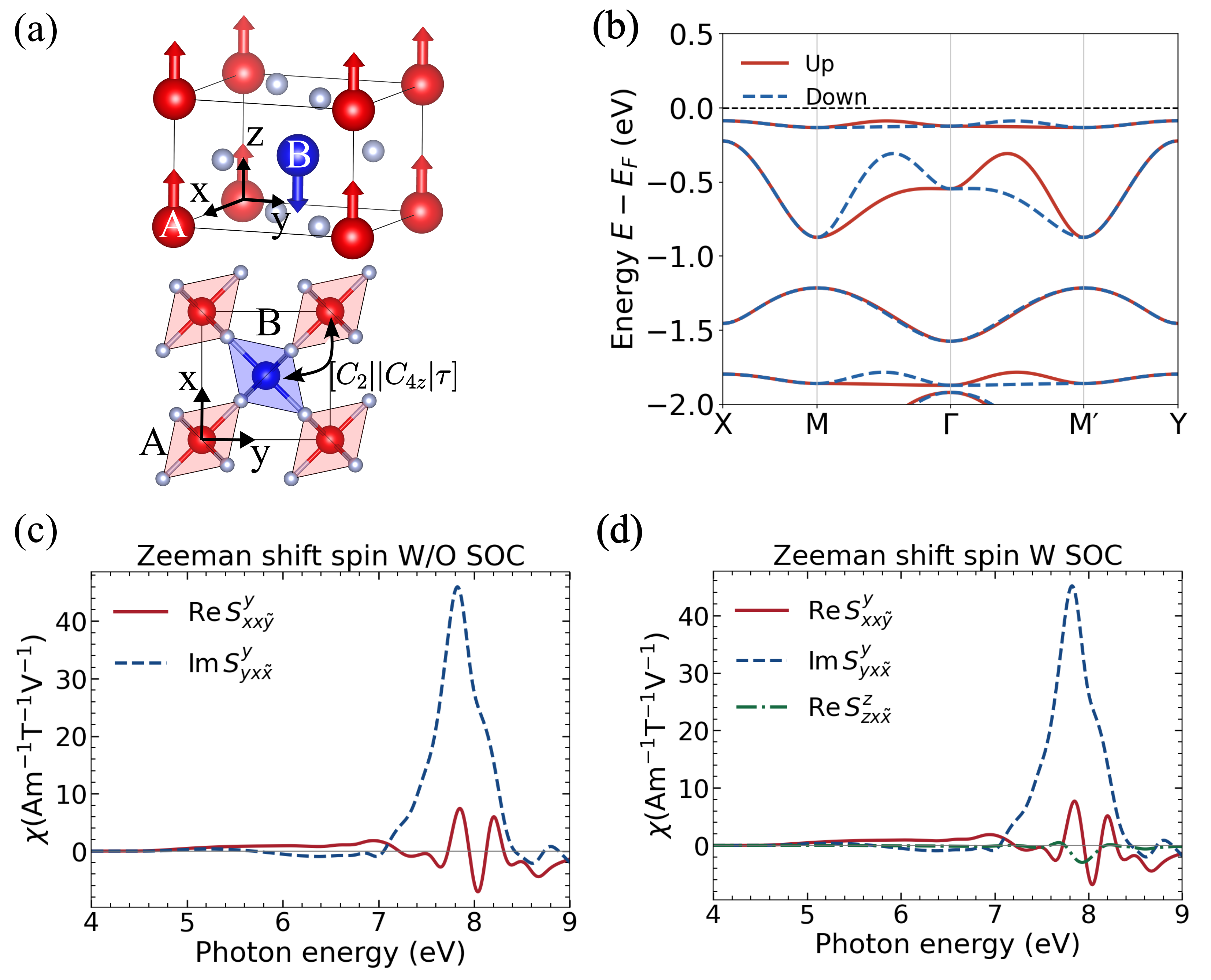} 
%width=8.6cm %scale=0.22
\caption{ 
(a) Crystal and magnetic structures of $\mathrm{MnF_2}$. Red and blue Mn sites carry opposite moments; gray spheres denote F atoms.
(b) Spin-resolved bands without SOC. Only valence bands are shown to highlight their pronounced spin splitting.
(c),(d) Zeeman spin-shift spectra without and with SOC, respectively, demonstrating a finite transverse spin-shift response even in the absence of SOC.
} 
\label{Fig3}
\end{figure}

\textcolor{blue}{\textbf{SOC-free Zeeman spin-shift currents in $\mathrm{MnF_2}$}}---
We next demonstrate SOC-free Zeeman spin photocurrents in the collinear antiferromagnet $\mathrm{MnF_2}$~\cite{Erickson1953,Yuan2020,Faure2026,Bhowal2024}.
Its rutile magnetic structure hosts antiparallel Mn moments along $c$ and belongs to the centrosymmetric spin point group ${}^{2}4/{}^{1}m{}^{2}m{}^{1}m$, as shown in Fig.~\ref{Fig3}(a).
Inversion symmetry forbids conventional $EE$ shift and injection currents of both charge and spin, while $EB$ Zeeman shift and injection charge currents vanish without SOC~\cite{Dong2026zeeman,Xiang2026}, favoring nearly pure spin photocurrents when SOC is weak.

The calculated band structure of $\mathrm{MnF_2}$ agrees with previous first-principles results~\cite{Yuan2020} (see Supplemental Material~\cite{supp,Kresse1996,perdew1996,Dudarev1998} for computational details). Figure~\ref{Fig3}(b) shows only the valence bands to highlight their spin splitting. We consider light propagating along $z$, with its electric and magnetic fields in the $xy$ plane. Fig.~\ref{Fig3}(c) shows the linear and circular Zeeman spin-shift responses $\operatorname{Re}S^y_{xx\tilde{y}}$ and $\operatorname{Im}S^y_{yx\tilde{x}}$, respectively. Both remain finite without SOC and carry spin polarization transverse to the N\'eel vector. With charge photocurrents and spin injection absent in this limit, the response is a pure spin-shift current.
 
Although SOC is present in real materials, it leaves the dominant spin-shift responses nearly unchanged. With SOC included, the response is constrained by the magnetic point group $4'/mmm'$, and additional components, such as the linear response $\operatorname{Re}S^z_{zx\tilde{x}}$ in Fig.~\ref{Fig3}(d), become finite but remain small owing to the weak SOC in $\mathrm{MnF_2}$.

Using a conservative electron spin $g$ factor of $g=2$, we obtain a bulk Zeeman spin-shift photoconductivity reaching several tens of $\mathrm{A/(mTV)}$. Under identical illumination conditions, the resulting current is approximately two orders of magnitude smaller than conventional shift currents in noncentrosymmetric materials and is independent of the carrier lifetime. Reported observations of conventional shift currents under weak illumination, including simulated sunlight~\cite{Glass1974,Young2012,Nakamura2017,Dai2023,Chan2021}, suggest that Zeeman spin-shift currents may also be experimentally accessible.

\begin{comment}
To assess experimental feasibility, we consider a representative optical electric-field amplitude $E=10^4~\mathrm{V/m}$, corresponding to $B=E/c\simeq3\times10^{-5}~\mathrm{T}$ for propagating light.
The estimated Zeeman injection and shift current densities reach $10^{2}$--$10^{3}~\mathrm{A/m^{2}}$ and \(1\)–\(10~\mathrm{A/m^{2}}\), respectively. For comparison, conventional shift-current densities in noncentrosymmetric materials typically reach \(10^{2}\)–\(10^{3}~\mathrm{A/m^{2}}\) at the same electric-field amplitude. The reported observation of conventional shift currents under weak illumination~\cite{Choi2009,Grinberg2013,Hatada2020,Nakamura2017}, including simulated sunlight, suggests that both Zeeman responses should be experimentally accessible. Owing to its larger magnitude, the Zeeman injection current should be more readily detectable than the Zeeman shift current.
\end{comment} 
 
In summary, we establish microscopic mechanisms for Zeeman spin-shift and spin-injection currents driven by interference between the electric-dipole and spin Zeeman couplings of light. These $EB$ responses enable spin-current generation in centrosymmetric magnets, where conventional $EE$ shift and injection currents are forbidden, making them particularly relevant to many prominent altermagnets that preserve inversion symmetry. We uncover a spin-shift current that survives without SOC and is independent of the carrier scattering time, and trace its origin to two spin flips mediated by the transverse Zeeman and spin-current vertices. Model calculations and first-principles results for $\mathrm{MnF_2}$ and $MnTe$ validate these predictions and indicate experimentally accessible response magnitudes. These findings provide a route to pure spin-current generation in centrosymmetric magnets, with potential applications in magnetic-state readout and polarization-sensitive photodetection.

\begin{acknowledgments}
This work was supported by the National Natural Science 
Foundation of China (Grants No.~12674287)
and the National Key R\&D Program of China (Grant No.~2025YFA1411200).
\end{acknowledgments}

%\textit{Data availability}---Details of the model and numerical calculations are provided in the main text and Supplemental Material. The source code for the model calculations is publicly available~\cite{data}.

%\textit{Data availability}--The data that support the findings of this study were generated by numerical simulations. The source code used to generate the simulations is publicly available~\cite{data}.

\nocite{showArticleTitles}
\bibliographystyle{apsrev4-2}
\bibliography{optical}

%\appendix
%\section{Appendix Title}
%%%%%%%%%%%%%%%%%%%%%%%%%%%%%%%%%%%%%%%%
% Supplemental Material
%%%%%%%%%%%%%%%%%%%%%%%%%%%%%%%%%%%%%%%%

%%%%%%%%%%%%%%%%%%%%%%%%%%%%%%%%%%%%%%%%
% Supplement References
%%%%%%%%%%%%%%%%%%%%%%%%%%%%%%%%%%%%%%%%

%%%%%%%%%%%%%%%%%%%%%%%%%%%%%%%%%%%%%%%%
% Supplement References
%%%%%%%%%%%%%%%%%%%%%%%%%%%%%%%%%%%%%%%%

%\bibliographystyle{apsrev4-2}
%\bibliography{suppref}

\end{document}